\documentclass[aaspp4,11pt,preprint,apjfonts]{emulateapj}

\slugcomment{accepted for publication in ApJ}
\shorttitle{SIMPLE star formation in the E-CDFS}
\shortauthors{Damen et al.}

\newcommand{\lum}{luminosity}
\newcommand{\lums}{luminosities}

\newcommand{\spitzer}{\textit{Spitzer}}

\newcommand{\Mstar}{\hbox{$M_*$}}
\newcommand{\Msol}{\hbox{$M_\odot$}}

\newcommand{\msol}{\hbox{$M_\odot$}}

\newcommand{\lsol}{\hbox{$L_\odot$}}
\newcommand{\infinity}{\hbox{$\infty$}}

\newcommand{\um}{\hbox{$\mu$m}}

\newcommand{\mone}{\hbox{$[3.6\mu\mathrm{m}]$}}
\newcommand{\mtwo}{\hbox{$[4.5\mu\mathrm{m}]$}}
\newcommand{\mthree}{\hbox{$[5.8\mu\mathrm{m}]$}}
\newcommand{\mfour}{\hbox{$[8.0\mu\mathrm{m}]$}}

\def\figa{
  \begin{figure}
    \includegraphics[width=0.5\textwidth]{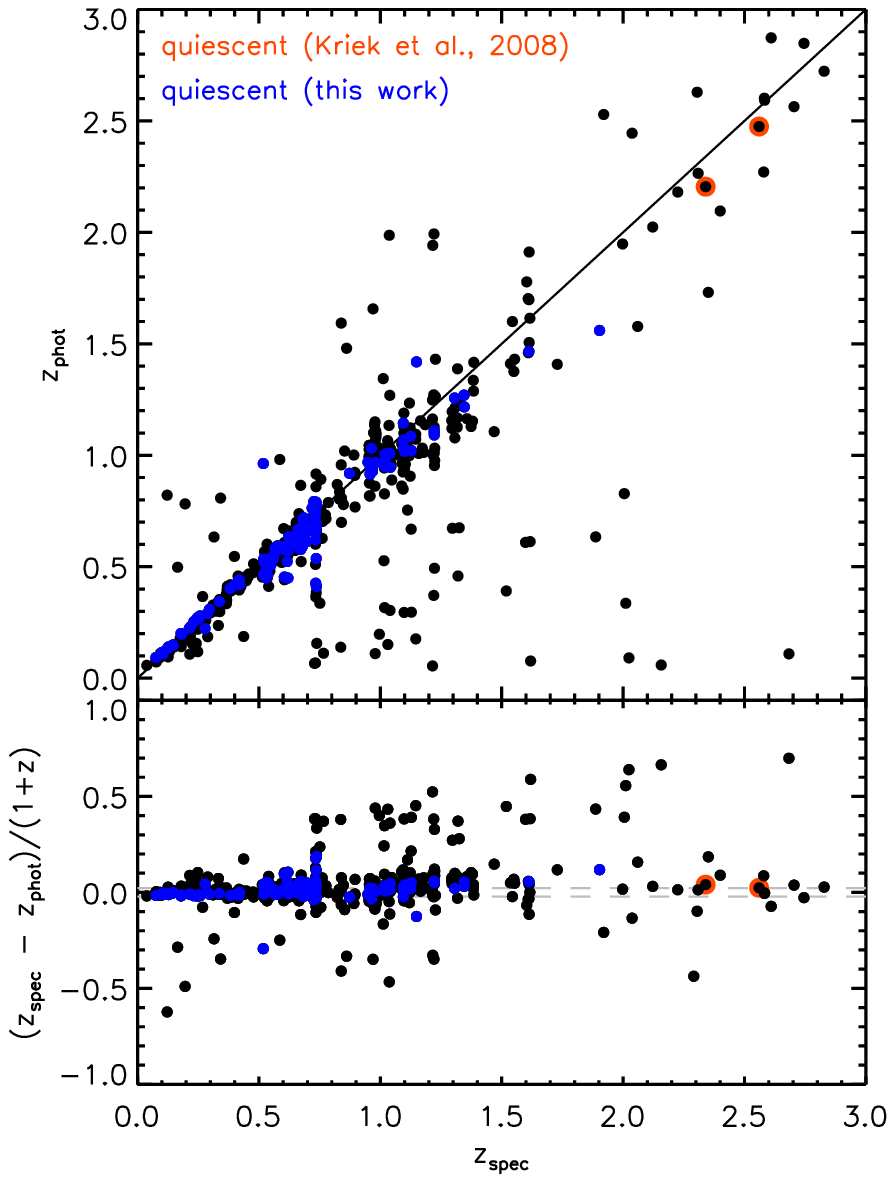}
    \caption[specz.eps]{The estimated $z_{phot}$ and $ dz = (z_{spec} -z_{phot}) / ( 1 + z_{spec})$ versus $z_{spec}$. Photometric redshifts are taken from the COMBO-17 survey \citep{wolf} or determined using EAZY, a photometric redshift code that uses a linear combination of templates to find the best redshift. The lower panel shows $dz$. Despite the dramatic outliers, the median absolute value of $|dz| = 0.033$, which is represented by the gray dashed line. Overplotted in blue are the galaxies that we define to be quiescent (see section \ref{res} for the applied criterion). Shown in red are overlapping quiescent galaxies from the sample of \citet{kriek08}.
      \label{specz}}
  \end{figure}
}

\def\figb{
  \begin{figure}
    \includegraphics[width=0.4\textwidth]{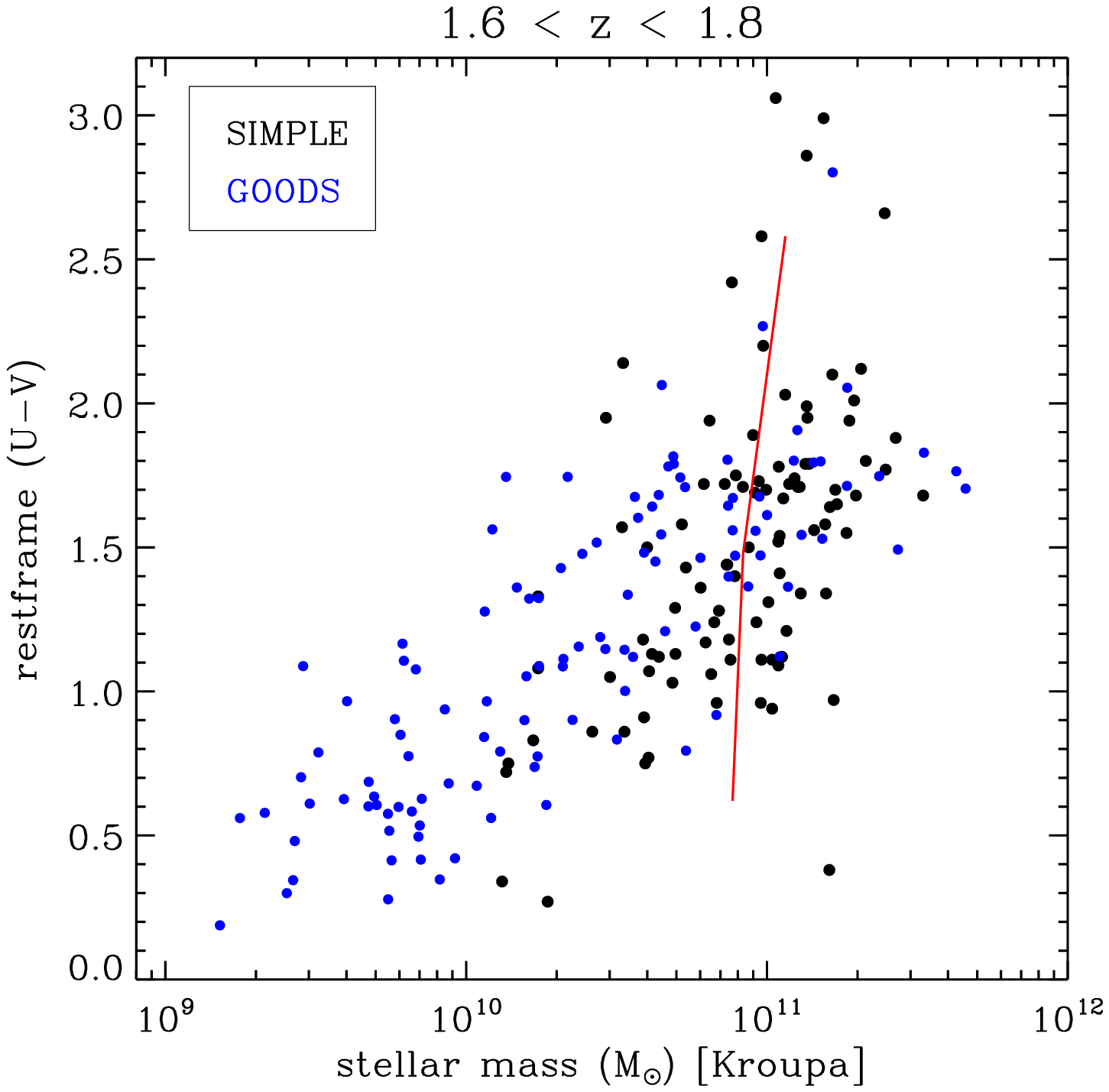}
    \caption[complete.eps]{Mass completeness: Rest-frame U-V color versus stellar mass for $1.6 < z < 1.8$. Our sample is shown in black dots. The red line is determined by scaling the detected sources down to the SIMPLE detection limit. It shows the minimal mass for 90\% of these scaled down sources, which means that out to $z\sim1.8$, we are complete for galaxies with $ M_{*}> 10^{11}\Msol$. Blue points refer to the deeper GOODS data, added to illustrate the incompleteness at the low-mass end.
       \label{mass}}
  \end{figure}
}

\def\figc{
  \begin{figure}
    \begin{center}
    \includegraphics[width=0.5\textwidth]{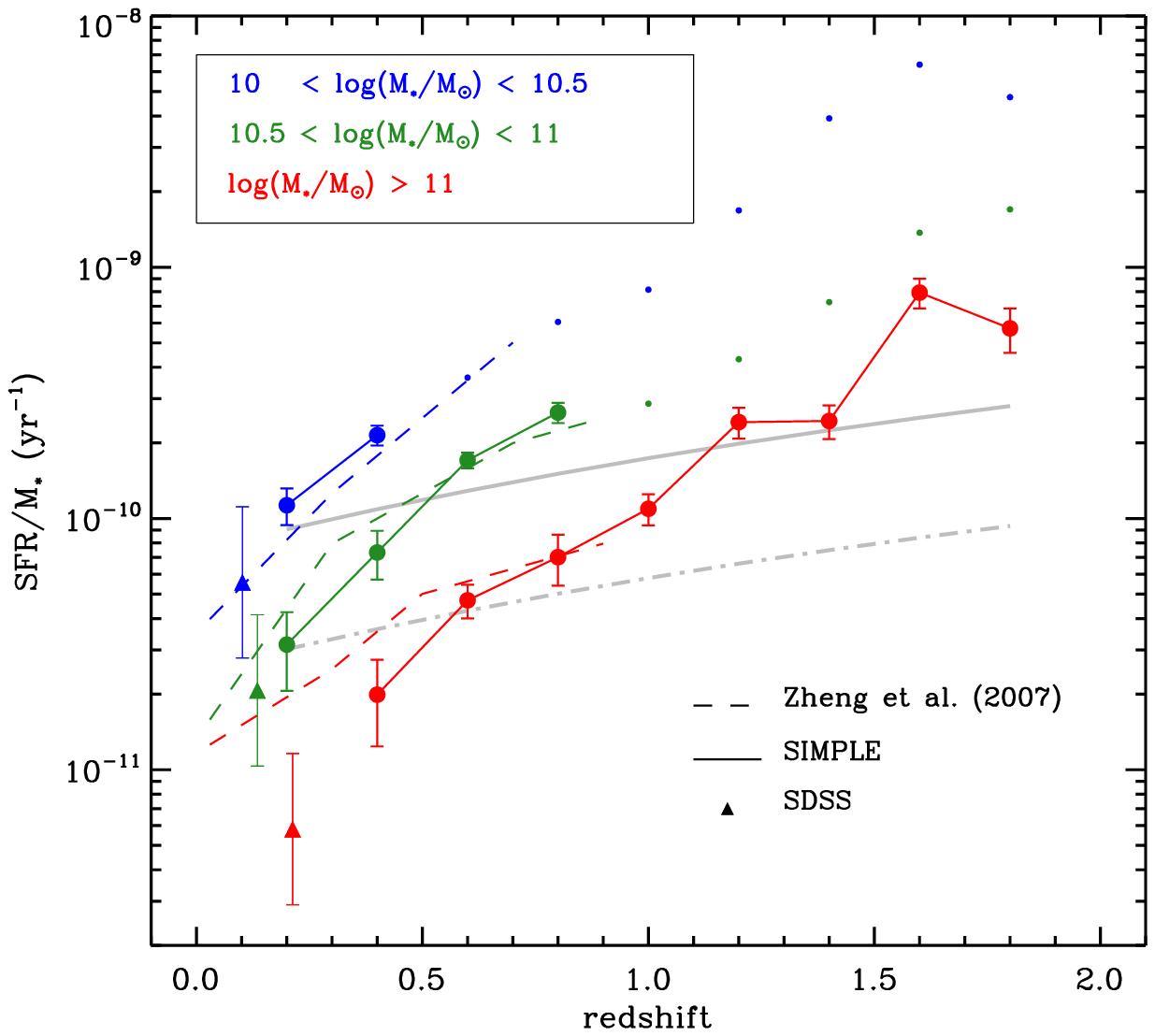}
    \caption[ssfr_z.eps]{SSFR versus redshift in different mass bins. Filled circles are SIMPLE results, dots show where we become incomplete with respect to mass. Triangles denote SDSS data. The error bars represent bootstrap errors for SIMPLE and a systematic error of 0.3 dex for the SDSS data. The dashed colored lines represent the results from \citet{zheng} in identical mass bins. The gray solid line is the inverse of the Hubble time ($1/t_{H}$ in $yr^{-1}$). Sources above this line are in a starburst mode: the time they needed for their stars to form is shorter than the Hubble time. Star formation is quenched in galaxies under the gray dashed line ($1/(3\times t_{H})$); the bulk of their stars has already been formed. The SSFR increases with $z$ at a rate that appears independent of  mass and SSFRs of more massive galaxies are typically lower than those of less massive galaxies over the whole redshift range. These results both confirm and expand the findings of \citet{zheng}. \\
      \label{ssfr}}
    \end{center}
  \end{figure}
}

\def\figd{
  \begin{figure}
    \begin{center}
      \includegraphics[width=0.5\textwidth]{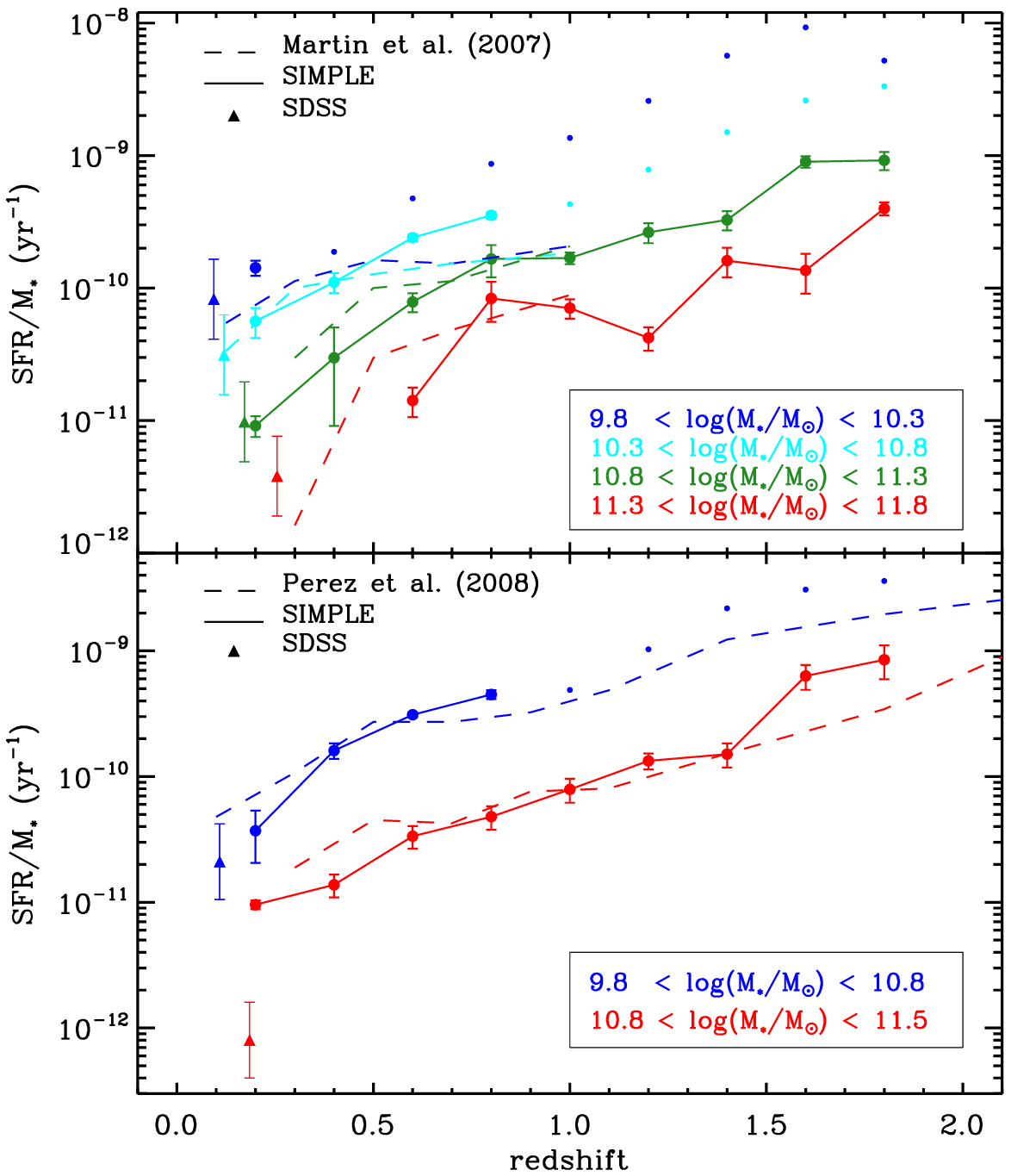}
      \caption[ssfr_z2.eps]{Same as Fig.~\ref{ssfr}, now compared with results from Martin et al. (2007) (upper panel) and Perez-Gonzalez et al. (2008) (lower panel). The mass bins differ from those in Fig.~\ref{ssfr} and were determined by subtracting a factor of 0.2 dex from the mass bins the quoted authors use, to correct for the difference in IMF. SSFR values in the lower pannel are calculated using the median, following Perez-Gonzalez et al. (2008).\\
        \label{ssfr_comp}}
    \end{center}
  \end{figure}
}

\def\fige{
  \begin{figure*}
    \includegraphics[width=\textwidth]{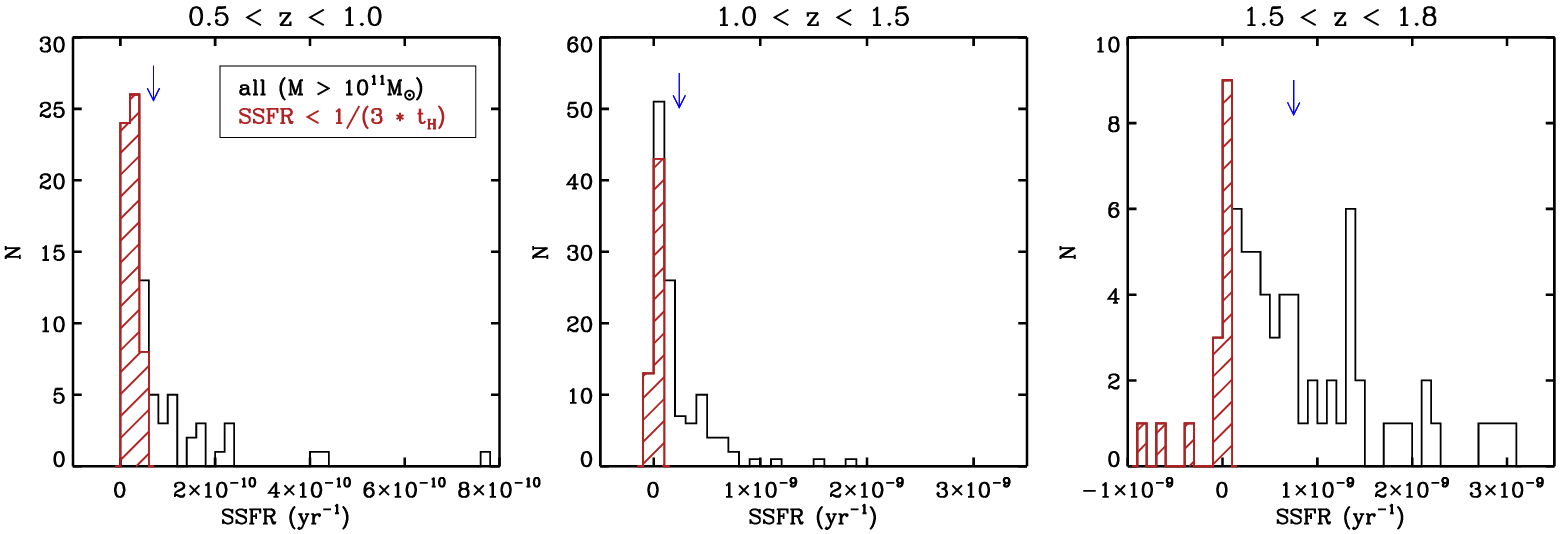}
    \caption[ssfrhist.eps]{The distribution of the SSFR for galaxies with $\Mstar > 10^{11} \msol$ in three redshift bins: left: $0.5 < z < 1.0$, middle: $1.0 < z < 1.5$, right: $1.5 < z < 1.8$. In all three redshift regimes the distribution is quite wide and peaks at low SSFRs. Blue arrows point to the average value of the SSFR in each redshift bin.
      \label{hist}}
  \end{figure*}
}

\def\figf{
  \begin{figure}
    \includegraphics[width=\columnwidth]{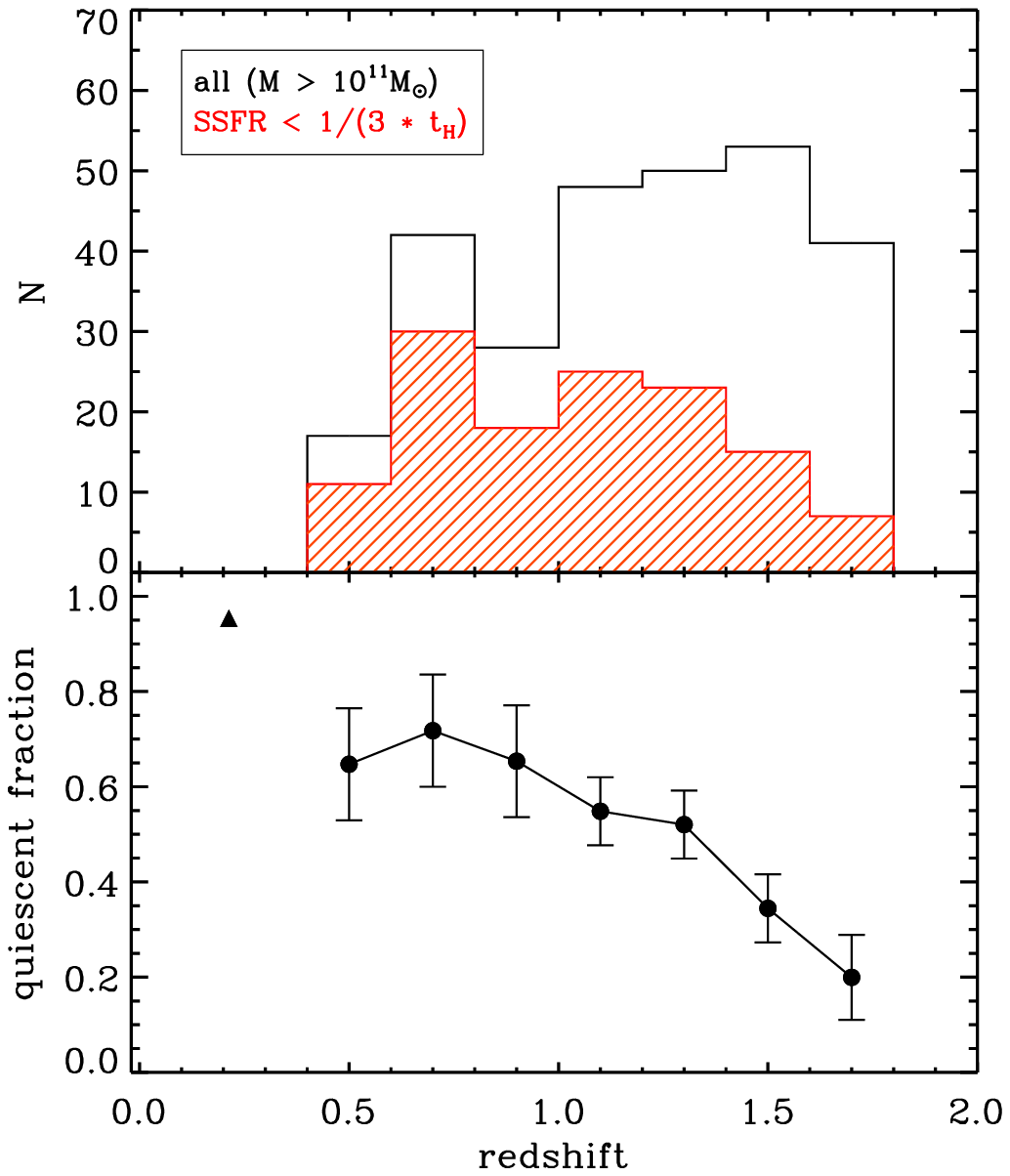}
    \caption[fraction.eps]{Overview of the fraction of quiescent galaxies in the highest mass bin ($ M_{*}> 10^{11}\Msol$), for which we are complete out to $z \sim 1.8$. Quiescent galaxies are defined as sources with SSFR $< 1/(3\times t_{H})\, yr^{-1}$, where $t_{H}$ is the age of the universe at a given redshift. The upper panel shows a histogram of all galaxies in this mass range. Overplotted in red is the number of galaxies whose star formation is quenched. The lower panel shows the fraction of galaxies in quiescent mode, determined from the histogram values. The error bars represent bootstrap errors. SDSS data have been used to determine a local value (triangle).
      \label{frac}}
  \end{figure}
}

\begin{document}

\title{The evolution of the specific star formation rate of massive galaxies to $z \sim 1.8$ in the E-CDFS}

\author{Maaike Damen\altaffilmark{1}, Ivo Labb\'e\altaffilmark{2}, Marijn Franx\altaffilmark{1}, Pieter G. van Dokkum\altaffilmark{3}, Edward N. Taylor\altaffilmark{1}, Eric J. Gawiser\altaffilmark{4}}

\email{damen@strw.leidenuniv.nl}

\altaffiltext{1}{Leiden Observatory, Leiden University, PO Box 9513,
  2300 RA Leiden, The Netherlands}
\altaffiltext{2}{Carnegie Observatories, 813 Santa Barbara Street, Pasadena, CA 91101; Hubble Fellow}
\altaffiltext{3}{Department of Astronomy, Yale University, New Haven, CT, 06520}
\altaffiltext{4}{Department of Physics and Astronomy, Rutgers University, Piscataway, NJ 08854}

\begin{abstract} 
  We study the evolution of the star formation rate (SFR) of mid-infrared (IR)
selected galaxies in the extended Chandra Deep Field South (E-CDFS). We use a combination of U-K GaBoDS and MUSYC data, deep IRAC observations from SIMPLE, and deep MIPS data from FIDEL. This unique multi-wavelength data set allows us to investigate the SFR history of massive galaxies out to redshift $z \sim 1.8$. We determine star formation rates using both the rest-frame ultraviolet luminosity from young, hot stars and the total IR luminosity of obscured star formation obtained from the MIPS $24 \,\um$ flux. We find that at all redshifts the galaxies with higher masses have substantially lower specific star formation rates than lower mass galaxies. The average specific star formation rates increase with redshift, and the rate of incline is similar for all galaxies (roughly $(1+z)^{n}, n = 5.0 \pm 0.4$). It does not seem to be a strong function of galaxy mass. Using a subsample of galaxies with masses $M_{*}> 10^{11}\Msol$, we measured the fraction of galaxies whose star formation is quenched. We consider a galaxy to be in quiescent mode when its specific star formation rate does not exceed $1/(3\times t_{H})$, where $t_{H}$ is the Hubble time. The fraction of quiescent galaxies defined as such decreases with redshift out to $z \sim 1.8$. We find that, at that redshift, $19 \, \pm9\%$ of the $M_{*}> 10^{11}\Msol$ sources would be considered quiescent according to our criterion. 
\end{abstract}


\section{Introduction} 
The star formation history of massive galaxies is not well understood.  Standard galaxy formation models have difficulty reproducing today's red and dead galaxies, 
unless mechanisms are introduced that prevent the gas from cooling and forming stars. 
To better constrain the models, it is useful to determine the star formation rates (SFRs) of galaxies as a function of mass and redshift. This has been done out to redshift $z = 1$, using the COMBO-17 survey \citep{zheng}.\\
 A surprising result of their study was that the specific star formation rates (SFR per unit stellar mass, SSFR) of high mass galaxies evolve at the same rate as those of low mass galaxies, where the most massive galaxies are offset to lower SSFRs. At higher redshifts, studies of star formation have so far focused mainly on either specific galaxy populations or specific redshift regimes (e.g, Lyman break galaxies, (Steidel et al. 1996, 1999), distant red galaxies,  (Papovich et al. 2006)). Papovich et al. (2006) found that massive, red galaxies at $1.5 \le z \le 3$ have SSFRs that are comparable to the global value integrated over all galaxies. Given the fact that we can already see the Hubble sequence in place at $z \sim 1$, this means that the period between $1 \lesssim z \lesssim 2.5$ is an interesting stage of transition, where massive galaxies evolve from actively star forming systems to the passive galaxies we observe in the local universe. The connection between the high and low redshift galaxy populations is not yet clear. \\
In this paper, we investigate the star formation history of massive galaxies ($\Mstar > 10^{11} \msol$), through measurements of the specific star formation rate from $z \sim 0.2$ to $z \sim 1.8$. We use a combination of a new, wide field \spitzer/IRAC survey called SIMPLE and ancillary data ranging from the near-ultraviolet (near-UV) to the mid-infrared (MIR). \\
Throughout the paper we assume a $\Lambda$CDM cosmology with $\Omega_{\rm m}=0.3$, $\Omega_{\rm  \Lambda}=0.7$, and $H_{\rm 0}=70$~km s$^{-1}$ Mpc$^{-1}$. All magnitudes are given in the AB photometric system. We denote magnitudes from the four \spitzer\ IRAC channels as \mone, \mtwo, \mthree, and \mfour, respectively. Stellar masses are determined assuming a Kroupa initial mass function (IMF).

\section{Data}\label{dat}
\subsection{Observations and sample selection}\label{sam}
We have combined imaging from the near-UV to MIR for this paper. The IR imaging was primarily taken from the SIMPLE survey (\spitzer's IRAC and MUSYC Public Legacy of the Extended CDFS). This survey consists of deep observations with the Infrared Array Camera (IRAC; Fazio et al. 2004) covering the 0.5 x 0.5 deg area centered on the Chandra Deep Field South (CDFS) in wavelength bands 3.6 \um, 4.5 \um, 5.8 \um \,, and 8.0 \um. The SIMPLE IRAC observations are supplemented with the IRAC images from the Great Observatories Origins Deep Survey (GOODS; Dickinson et al. (in prep.) ). These very deep images were taken on the central $\sim \,160\,$ arcmin$^2$ \, of the field. The combined mosaics are publicly available from the \spitzer \, Science Center\footnote{http://data.spitzer.caltech.edu/popular/simple}. A detailed description of the observations and data reduction will be given in Damen et al. (in prep.). \\
For coverage of the optical/near-infrared (NIR) wavelengths, we used ground-based data from different sources. To cover the UV to optical regime, we used the UBVRI imaging from the COMBO-17 and ESO DPS surveys (Wolf et al. 2004 and Arnouts et al. 2001, respectively) in the re-reduced version of the GaBoDS consortium (Erben et al. 2005, Hildebrandt et al. 2006). We obtained $z$'JHK images from the Multiwavelength Survey by Yale-Chile (MUSYC, Gawiser et al. 2006), which are available online\footnote{http://www.astro.yale.edu/MUSYC}. The final UBVRI$z$'JHK images we used were position-matched by Taylor et al. (in prep.). We also include the MIR 24 \um\, MIPS image from the Far-Infrared Deep Extragalactic Legacy Survey (FIDEL, Dickinson et al. (in prep.)). \\
Sources were detected and extracted using the SExtractor software (Bertin \& Arnouts 1996) on a detection image, which is an inverse-variance weighted average of the most sensitive IRAC bands, 3.6 and 4.5 \um. The images were convolved with a Gaussian to match the 8.0 \um\, image, which has the broadest FWHM ($\sim 2.0''$). Using an aperture diameter of $4"$, we detected $\sim$ 61,000 galaxies to a limiting depth of $ (\mone + \mtwo)/2 < 24.0$ (24.3 for the GOODS area). \\
By selecting all sources with  $ (\mone + \mtwo)/2 < 21.2$, we created a  subsample of 3841 sources, 95\% of which have $S/N > 5$ in K. From this subsample we excluded all X-ray detected sources as they are highly likely active galactic nuclei (AGN; Virani et al. 2006 and Alexander et al. 2003). Stars were identified using the color criterion $J - K < 0.04$ and removed from the sample. The final sample contains 3393 sources out to $z = 2$. From this sample, 60\% of the sources are detected in MIPS ($S/N > 10$). At $z \sim 1.8$, our highest redshift bin, 83 \% of the sources are detected in MIPS. Such high detection rates are consistent with earlier results (Daddi et al. 2005, Papovich et al. 2006). Since we interpret 24 \um \, flux directly as star formation activity (rather than evidence of AGN activity), the high fraction of MIPS detected sources contributes greatly to our conclusions regarding the star formation history (see sections \ref{sfr} and \ref{conc}).\\
The MIPS fluxes in particular were treated for blending. We used the IRAC 3.6 \um \, image, which has a smaller PSF to subtract modeled sources from MIPS sources that showed close neighbors, thus deblending the image (see Labb\'e et al. (2004, 2006) for more information on this technique). For the IRAC images themselves, which also suffer from blending, this method could not be applied since the K-band image we would like to use for this is not deep enough for this kind of modeling. We compared our final MIPS fluxes with the deeper observations of the GOODS team as a check and found that at the faint end, our fluxes were slightly larger. This could be an effect of remaining blending issues and we investigate this further and see how it affects our results in section \ref{conc}.

\figa

\subsection{Spectroscopic and photometric redshifts}\label{ez}
The E-CDFS has been intensely targeted for observations the last few years and, as a result, many spectroscopic redshifts are available for our sample. We collected 438 spectroscopic redshifts from large surveys by Cimatti et al. (2002), le F\`evre et al. (2004), Vanzella et al. (2008), and Ravikumar et al. (2007), which accounts for 13 \% of our sample. In addition, we included photometric redshifts from the COMBO-17 survey out to $z = 0.7$ \citep{wolf}. For the remainder of the sources we used the new photometric redshift code EAZY (Brammer et al. 2008) to obtain redshifts. \\
Figure~\ref{specz} shows the available spectroscopic redshifts versus photometric redshifts from COMBO-17 and EAZY. We measure the scatter by determining the median absolute deviation of $|dz| = 0.033$ where $dz = (z_{spec} -z_{phot}) / ( 1 + z_{spec})$. For $z \ge 1$, which is the regime we are specifically interested in, this value is somewhat higher: $|dz| = 0.079$.\\
In Section \ref{res} we will inspect the fraction of quiescent galaxies. Uncertain photometric redshifts can affect this fraction and it is, therefore, important to verify that for the quiescent galaxies the photometric redshifts are not dramatically offset. The blue dots in Fig.~\ref{specz} represent sources we classify as quiescent. Their photometric redshifts do not lie among the most extreme outliers and their median absolute deviation is $|dz| = 0.024$ (0.050 at $z \ge 1$), which is smaller than for the complete sample. To check the accuracy of our photometric redshifts at $z \ge 2$, we use the spectroscopic survey of \citet{kriek08}. We have only included the sources they classified as quiescent (red dots). The median offset for these sources is $|dz| = 0.059$.

\subsection{Low-redshift sample}
We include data from the Sloan Digital Sky Survey (SDSS) to check whether we are consistent with the local universe. SDSS masses were determined by Kauffmann et al. (2003) using spectra. Brinchmann et al. (2004) derived SFRs from emission lines. For details on the derivation of the masses and SFRs in the SDSS we refer to their papers.  

\section{Star formation rates, mass and completeness}
\subsection{Inferring the SFRs from the 24 \um\, flux and UV luminosity}\label{sfr}
We estimated SFRs using the UV and IR emission of the sample galaxies. The UV flux probes the unobscured light from young stars, whereas the IR flux measures obscured star formation through light that has been re-processed by dust. Combined they give a complete census of the bolometric \lum \, of young stars in the galaxy (Gordon et al. 2000, Bell 2003).\\
At the redshifts of interest ($z \sim 0.2-1.8$), MIPS 24 \um \, probes rest-frame 8-15 \um, which broadly correlates with the total IR \lum\, ($L_{IR} = L(8-1000 $\um). We use IR template spectral energy distributions (SEDs) of star forming galaxies of Dale \& Helou (2002) to translate the observed 24 \um \, flux to $L_{IR}$. First, we convert the observed 24 \um \, flux density to a rest-frame \lum \,density at $24/(1+z) $\,\um, then we extrapolate this value to a total IR \lum \,using the template SEDs. The model spectra cover a wide range of spectral shapes, allowing for different heating levels of the interstellar environment. Following \citet{wuyts} we adopt the mean of log($L_{IR}$) derived from the templates ranging from quiescent to active galaxies as the best estimate for the the total IR luminosity. To convert the UV and IR \lums \, to a SFR, we use the calibration from Bell et al. (2005), which is in accordance with \citet{pap06}, using a Kroupa IMF:

\begin{equation}\label{eqn:ir_sfr}
\Psi / \msol \,\,\mathrm{yr}^{-1} = 1.09 \times 10^{-10} \times
(L_\mathrm{IR} + 3.3\,\, L_{2800}) / \lsol,
\end{equation}\\

where $L_{2800} = \nu L_{\nu,(2800 \AA)}$ is the \lum \, at rest-frame 2800 \AA, a rough estimate of the total integrated UV \lum \,(1216-3000\AA). 
The scatter in the conversion to $L_{IR}$ induces a systematic error of typically 0.3 dex (Bell et al. 2005, Papovich et al. 2006). Another source of error is the uncertainty in photometric redshifts, as small changes in redshift can have a significant effect to the conversion. Applying the 68\% confidence values of the photometric redshifts induces variations in the inferred $L_{IR}$ of 0.1 dex.\\
There are some additional sources of error that are harder to quantify. Firstly there is the assumption that local IR SEDs represent the high-redshift galaxy population accurately. The reliability of this assumption has been investigated by \citet{adel}, who found that the bulk of intermediate to high-redshift galaxies have IR SEDs similar to galaxies in the local universe. However, the physical grounds for this are still unknown. Secondly, an AGN would also contribute to the 24 \um \,emission. Although we removed all X-ray detections from our sample, dust-obscured AGN could still be present and some SFRs may in fact be upper limits.

\subsection{Stellar mass and rest-frame colors}

We fitted the UV-to-8 \um \, SEDs of the galaxies using the evolutionary synthesis code developed by \citet{bc03} to obtain stellar masses for our sample. We assumed solar metallicity, a Salpeter IMF and a Calzetti reddening law. We used the publicly available HYPERZ stellar population fitting code \citep{bol} and let it choose from three star formation histories: a single stellar population (SSP) without dust, a constant star formation (CSF) history and an exponentially declining star formation history with a characteristic timescale of 300 Myr ($\tau 300$), the latter two with varying amounts of dust. To facilitate comparison with other studies, the derived masses were subsequently converted to a Kroupa IMF by subtracting a factor of 0.2 dex. We calculated rest-frame luminosities and colors by interpolating between observed bands using the best-fit templates as a guide (see \citet{rud} for a detailed description of this approach).

\subsection{Mass completeness}
\figb
To determine the mass limit to which we are complete, we take detected sources with $ 1.6 < z < 1.8 $ and scale them down in mass to the flux detection limit ($ (\mone + \mtwo)/2 = 21.2$).  This is illustrated in Fig.~\ref{mass} where rest-frame U-V colors are plotted against mass. The red line is the mass limit to which we can detect 90\% of the scaled sources. The black dots in this figure represent sources in our sample. Sources from the significantly deeper GOODS-ISAAC catalog \citep{wuyts} are overplotted in blue to illustrate the effect of incompleteness. We can conclude that we are 90\% complete for $M_{*} > 10^{11} \Msol$ in the highest redshift bin ($1.6 < z < 1.8$).

\section{Star formation rates as a function of redshift}\label{res}
We determine the average SFR in different mass bins to examine the evolution of specific SFR with redshift out to $z\sim 1.8$. The average is based on the SFRs determined from the UV and MIPS fluxes, as described in section \ref{sfr}. Sources with no significant MIPS flux were also included in the average.
\figc
\figd
\fige
Figure~\ref{ssfr} shows the redshift evolution of the average SSFR in different mass bins (filled circles). Dots show where we suffer from incompleteness, the error bars represent the bootstrapped 68\% confidence levels on the measurement of the mean SSFR. 
The SSFRs of more massive galaxies are typically lower than those of less massive galaxies over the whole redshift range. In addition, Fig.~\ref{ssfr} clearly shows that not only does the average SSFR rise monotonically with redshift (roughly following $(1+z)^n, n=5.0 \pm 0.4$, over our complete redshift range); but the rate of the change in SSFR also seems to be equally strong for galaxies of different mass. Naturally, the strength of this claim is reduced by the incompleteness at the low-mass end.  \\
The trends in SSFR we find for each mass bin are consistent with local values from SDSS data, represented in Fig.~\ref{ssfr} with triangles. We account for the difference in SFR derivation by applying a systematic error of 0.3 dex (J. Brinchmann, private communication). \\
Our results directly expand and confirm the findings of \citet{zheng}, who carried out a similar study based on a R-band selected sample in the E-CDFS and Abell 901/902. Their results are included in Fig.~\ref{ssfr} as dashed lines. For galaxies with $M_{*} > 10^{11} \Msol$ we can extend the trend in SSFR they find to $z \sim 1.8$. At $z < 0.6 $ the results diverge because of the low number (7) of sources in that specific bin. \\
In Fig.~\ref{ssfr_comp} we compare our results with other studies in similar fields: a study by \citet{martin} (E-CDFS; upper panel) and by \citet{pg08} (CDFS, Hubble Deep Field North and the Lockman Hole Field; lower panel). We converted the mass intervals of both studies to our choice of IMF and recalculated our mean SSFRs appropriate for these mass intervals.
Our data broadly agree with the results of \citet{martin}, especially in the intermediate mass bin ($10^{10.8} <M_{*} < 10^{11.3}$). However, there are discrepancies at the high-mass end, where the evolution of the SSFR with redshift they find is stronger than what we find, and at the low-mass end, where it appears to be weaker. As a result, Martin et al. (2007) do see a mass-dependence in the evolution of the SSFR. The reason for these differences is not immediately clear, particularly given the fact that both studies use the same field. We note, however, that Martin et al. (2007) use a different method for determining star formation rates: they derive SFRs from the UV and correct for extinction using MIPS. Furthermore, the difference at the high mass end may be caused by the poor number statistics in the highest mass bin ($M_{*} > 10^{11} \Msol$). The number of galaxies in this mass bin is 4, 3, and 6 for redshifts $z \sim 0.6, 0.8$, and $1.0$, respectively, which is probably comparable to their sample. Hence the significance of the difference at these masses is small and no strong statement about the evolution is possible.\\
\citet{pg08} use an IRAC selected sample and their SFRs are determined using a combination of rest-frame UV and MIPS 24 \um\, flux, similar to what we do. The lower panel of Fig.~\ref{ssfr_comp} shows their results, which are based on the median of the SSFR in each mass and redshift bin. The agreement out to $z \sim 1.4$ is good, beyond that, our median values are somewhat larger. Since the results come from different fields, it could be that field-to-field variation plays a role. Note that we use the mean SFR, rather than the median, in our main analysis, in contrast to \citet{pg08}. The mean SFR is (in our sample) on average a factor of 1.8 higher than the median.\\ 
Also shown in Fig.~\ref{ssfr} is the inverse of the Hubble time ($t_{H}$, gray solid line). Sources above this line are forming stars rapidly: the time they needed for their stars to form is shorter than the Hubble time. Sources below the line have had a declining SFR. Massive galaxies ($M_{*} > 10^{11} \Msol$) have on average a specific star formation rate of $\sim 2\,\times 10^{-10} yr^{-1}$ at $z \sim 1.1$, which is consistent with having a constant SFR over $z = \infinity$ to $ z \sim 1.1$. \\
Even though the average SSFR increases rapidly with redshift, the spread in SSFR is very high at all redshifts, with a peak at very low SSFRs. This is explicitly shown in Fig.~\ref{hist} which contains histograms of the SSFR of the most massive galaxies ($M_{*} > 10^{11} \Msol$) in three different redshift bins. The blue arrows point at the average value of the SSFR in each redshift bin. To characterize this low SSFR peak, we define quiescent galaxies with the criterion $SSFR < 1 / (3\times t_H)$. These galaxies must have had strong quenching of their star formation. This criterion is represented in Fig.~\ref{ssfr} by the dashed gray line. In Fig.~\ref{hist} the dashed red histograms represent sources that obey this criterion. \\
\figf
Figure~\ref{frac} shows the evolution of the fraction of quiescent galaxies thus defined with time. The upper panel shows a histogram of all galaxies with $M_{*} > 10^{11}$ and those that have SSFR smaller than  $ 1 /(3\times t_{H})\,\, yr^{-1}$ are overplotted in red. The lower panel shows the fraction of galaxies with quenched star formation as a function of redshift. SDSS data have been used to determine a local value (triangle). The error bars are again bootstrap errors. The fraction of quiescent galaxies decreases monotonically with redshift from the local universe out to $z \sim 1.8$, with the exception of $z \sim 0.5$. The fraction of passive galaxies in our lowest redshift bin seems inconsistent with this trend, which could be due to the low number of galaxies at this redshift. \\
Another thing to note is the elevated number of galaxies in the $z \sim 0.7$ bin, which shows a slightly higher quiescent fraction than its neighboring bins. This is probably due to overdensities known to exist at this redshift in the E-CDFS (Gilli et al. 2003, Wolf et al. 2004). Such overdensities may harbour more passive galaxies, which can account for the high quiescent fraction. 
The same effect can be seen in Fig. 9 of Kaviraj et al. (2008). This figure shows the evolution of recent star formation with redshift based on UV to optical colors. At $z \sim 0.7-0.75$ there is less star formation than in the neighboring redshift bins, which agrees with what we find. \\
The main point to take from Fig.~\ref{frac} is that we can still see massive quiescent galaxies out to $z \sim 1.8$, where they make up $19 \pm 9 \%$ of all massive galaxies.

\section{Conclusions}\label{conc}
We investigate the star formation history of massive galaxies out to redshift $z \sim 1.8$, by analyzing specific star formation rates (SSFRs) of a sample of $\sim 3,400$ sources from SIMPLE, a survey that combines new \spitzer/IRAC observations of the E-CDFS with ancillary data ranging from the near-UV to the MIR. We find quiescent galaxies with masses higher than $10^{11} \msol$ out to the highest redshift probed, $z \sim 1.8$. At this redshift, they form $19 \pm9\%$ of the total number of massive galaxies. The SSFR is an increasing function with redshift (roughly $(1+z)^{n}, n = 5.0 \pm 0.4$) for galaxies in all mass bins. The mean SSFRs are smaller in high-mass galaxies than in low-mass galaxies at all redshifts. It is interesting to consider this result in the context of "downsizing", a term which generally is taken to imply that more massive galaxies formed their stars before less massive galaxies (Cowie et al. 1996). An increasing amount of observational evidence supports this idea, in particular through studies of the (specific) star formation rate (Juneau et al. 2005, P\'erez-Gonz\'alez et al. 2005, Caputi et al. 2006, Papovich et al. 2006, Reddy et al. 2006 and Noeske et al. 2007). Figure \ref{ssfr} shows that the SSFRs of massive galaxies are the lowest of the whole sample. This indicates that they have already formed the bulk of their stars and that active star formation has shifted to the galaxies that are less massive. Additional support comes from a theoretical perspective. \citet{guo} investigated the contribution of star formation to galaxy growth in the Millennium Simulation. They found that even out to $z \sim 4-5$ less massive galaxies are always growing faster than galaxies of higher stellar mass.\\
It is interesting to see that although we see a change in the locus of star formation (from massive to less massive systems), we do not find any mass-dependence of the {\em evolution} of the specific star formation rates. High-mass galaxies and low-mass galaxies appear to evolve at the same rate out to $z\sim 1.8$, although deeper data are necessary to reduce possible effects of incompleteness of the lowest mass bins at high redshift.\\
Next we compare our passive fraction in our highest redshift bin to observational results from the literature. A recent estimate can be found in work by Labb\'e et al. (in prep.), who found $35 \pm 7\%$. We also re-determined the quiescent fraction for Kriek et al. (2006) by analyzing the full sample presented in Kriek et al. (2008). We defined all galaxies without emission lines and $SSFR < 0.05~ Gyr^{-1}$ to be quiescent. Out of the 28 galaxies at redshift $z > 2, 36 \pm 9\% $ are quiescent according to this method, applying a bootstrap error. The values of Labb\'e et al. (in prep.) and Kriek et al. (2008) are consistent with our fraction within $2\sigma$. We note that our value is lower, but we emphasize that all studies use different definitions which can influence the result. We return to that below.\\
To investigate the difference in the estimates of the quiescent fractions further, we look at our MIPS fluxes and compare them with results from the CDFS (Labb\'e et al. 2005). For the overlapping sources in their sample and ours, we observed a median positive offset in MIPS flux of $4\, \mu Jy$ with respect to the CDFS, which means our SFRs are overestimated. To investigate whether this offset could be responsible for the difference in passive fraction, we simulated the effect errors in MIPS flux would have on our results. We randomly added a measurement of the difference in MIPS flux to a collection of simulated passive galaxies with $M_{*} > 10^{11} \Msol$ and determined the number of times the SSFR of such a source would scatter above the limit by which we define a passive galaxy. Fifteen percent of the passive galaxies were classified as star forming after performing this test. This raises the fraction of quiescent galaxies at redshift $z \sim 1.8$ by 2\%, which is not enough to explain the difference between the fractions.\\
We next compare our quiescent classification with the results from rest-frame optical spectroscopy of Kriek et al. (2008). 
Their sample contains 11 sources in the E-CDFS, 2 of which show no emission lines and are best fit with a passively evolving SED. 
We detect one of these sources in 24 \um \, with $S/N < 1$. Our SFR for it is $9 \pm 10 \,\Msol yr^{-1}$, which is consistent with the $0.7\pm 0.7 \,\Msol yr^{-1}$ that Kriek et al. (2008) find. The other source has a SSFR of $2.4 \times \,10^{-10}\,yr^{-1}$, which exceeds the limit of $1 / (3\times t_H)$. It is still likely to be quenched as its SSFR is smaller than $1 / t_H$. In summary, our results agree reasonably well for these two sources, but the exact definition of quiescence may cause variations in the result. 
As to illustrate this further, we relaxed our limit to $SSFR < 1 / t_H$. For this limit, we find a quiescent fraction of $30\pm 7\%$ at $z \sim 1.8$ which is roughly 1.5 times the fraction we found earlier.\\
In addition, the 24 \um \,emission we detect could be due to the presence of a weak and obscured AGN. This would mean that some of the galaxies we call star forming could in fact be quiescent galaxies hosting an obscured AGN. If this is the case, our fraction underestimates the real fraction. This is possible as evidence exists that AGN activity is widespread among massive galaxies at these redshifts (Rubin et al. 2004, Daddi et al. 2007,  Kriek et al. 2007). We removed all X-ray detected sources from our sample as probable AGN candidates, but have no means to identify obscured AGNs that show strong 24 \um \,flux and weak X-ray emission. Our fraction would also be underestimated if the galaxy would hide an obscured starburst in its center.\\
Another effect is the error in the photometric redshifts, which we could have undervalued. The EAZY redshifts we use here are the results of several runs where we varied templates and error determination, which did not largely affect the outcome. We found that taking the 68\% confidence range on the photometric redshifts of the galaxies leads to variations in the inferred SFR of 0.1 dex, which is not enough to significantly affect the results.
Finally, there is a large diversity in the fields used for these studies and field-to-field variations could also be causing discrepancies. \\
Our most robust result is that we find a high fraction of galaxies with MIPS detections at redshift $z \sim 1.8$ and a small, but non-negligible fraction of quiescent galaxies, which we interpret as a lower limit. 
The galaxies that are detected in MIPS at redshift $z \sim 1.8$ are in some way active, either through star formation or black hole growth. Deeper 24 \um\, data and spectroscopic information will be crucial to be able to elaborate on this more.\\ \\

We thank the referee for useful suggestions. We thank the FIDEL team for early access to their 24 \um \,image and  in particular Jackie Monkiewicz, for reducing the 24 \um \, data. We are grateful to Jarle Brinchmann for his advice regarding the SDSS data and Natascha F\"orster-Schreiber for help with the SED fitting. This research was supported by grants from the Netherlands Foundation
for Research (NWO), and the Leids Kerkhoven-Bosscha Fonds. Support from National
Science Foundation grant NSF CAREER AST-0449678 is gratefully acknowledged.

\end{document}